\documentclass[pdflatex,sn-mathphys-num]{sn-jnl}

\usepackage{graphicx}%
\usepackage{multirow}%
\usepackage{amsmath,amssymb,amsfonts}%
\usepackage{amsthm}%
\usepackage{mathrsfs}%
\usepackage[title]{appendix}%
\usepackage{xcolor}%
\usepackage{textcomp}%
\usepackage{manyfoot}%
\usepackage{booktabs}%
\usepackage{algorithm}%
\usepackage{algorithmicx}%
\usepackage{algpseudocode}%
\usepackage{listings}%

\begin{document}

\title[Strong-Field QED at LUXE and Future Colliders]{From LUXE to Future Colliders: Probing Strong-Field QED and Beyond}


\author*[1,2]{\fnm{Ivo} \sur{Schulthess}}\email{ivo.schulthess@ethz.ch}

\affil[1]{\orgdiv{Institute for Particle Physics and Astrophysics}, \orgname{ETH Zurich}, \postcode{8093} \orgaddress{\city{Zurich}, \country{Switzerland}}}

\affil[2]{\orgname{Deutsches Elektronen-Synchrotron DESY}, \orgaddress{\postcode{22603} \city{Hamburg}, \country{Germany}}}


\abstract{Strong-field quantum electrodynamics offers a unique window into non-perturbative phenomena such as vacuum pair production, in which electron--positron pairs are created from the vacuum in the presence of intense electromagnetic fields. The LUXE experiment at DESY is designed to probe this regime using collisions between a high-intensity laser and the 16.5~GeV electron beam of the European XFEL. Future accelerator infrastructures, such as linear colliders, could extend these studies to even higher intensity and energy scales. Additionally, high-energy photons produced in such interactions can be used in beam-dump experiments to search for new physics.}



\maketitle

\section{Introduction and Motivation}

Quantum electrodynamics (QED) is among the most precisely tested theories in physics. Nevertheless, its behavior in the presence of electromagnetic fields approaching the critical QED scale or Schwinger field remains experimentally largely unexplored. In this strong-field QED regime, non-linear effects become significant and can become non-perturbative for sufficiently large field strengths. This gives rise to phenomena such as non-linear Compton scattering and non-linear Breit--Wheeler pair production. The theoretical framework of strong-field QED is reviewed in classic and modern treatments~\cite{Ritus:1985vta, Fedotov:2022ely, Kropf:2025loq}. 

While the critical field strength is far beyond what can be generated in the laboratory in a static configuration, it can be accessed effectively in the rest frame of ultra-relativistic particles. In electron–laser collisions, a large Lorentz factor of the order $10^4$ enhances the field strength seen by the electron, enabling experimental access to field strengths at or even beyond the Schwinger field at accelerator facilities. The interaction dynamics are determined by the beam energy and laser intensity, which act as experimentally adjustable quantities directly mapped onto the theoretical energy and intensity parameters of strong-field QED. 

The relevance of strong-field QED extends well beyond fundamental theory. Strong electromagnetic fields play an important role in a variety of physical systems, ranging from laser–plasma accelerators, crystal channeling, and beam–beam interactions at high-energy colliders operating at ever higher energies and intensities. In addition, strong-field QED processes provide controlled laboratory analogs of extreme astrophysical environments such as magnetars and black holes. 

These considerations motivate dedicated experimental efforts to explore SFQED in a controlled setting. The LUXE experiment at DESY represents the first precision experiment specifically designed to probe this regime using electron–laser collisions~\cite{Abramowicz:2021zja, LUXE:2023crk}. Beyond its role as a fundamental QED experiment, LUXE serves as a pathfinder for studies at future collider facilities by providing experimental input on strong-field effects in accelerator environments and by benchmarking theoretical and numerical descriptions of strong-field interactions. At the same time, the intense photon beams produced in these interactions enable novel opportunities for photon-based searches for new physics using fixed-target techniques.

\section{Strong-Field QED}

\subsection{The LUXE Experiment}

The LUXE experiment is designed to investigate strong-field QED in collisions between the 16.5~GeV electron beam of the European XFEL and a high-intensity optical laser. A staged approach using a laser with an increasingly high peak power of up to several hundred terawatts is foreseen. This strategy will allow systematic exploration of the transition from perturbative to non-perturbative QED dynamics.

At the interaction point, electrons, positrons, and photons can be produced. A dipole magnet placed downstream of the interaction region will spatially separate the charged particles from the photon beam, allowing each particle species to be measured independently. The experimental setup therefore enables a comprehensive study of the final-state particles emerging from strong-field interactions within a single apparatus.

Two processes form the core of the LUXE physics program. The first is non-linear Compton scattering, in which an electron emits a photon while interacting with multiple laser photons. In contrast to linear Compton scattering, this process leads to characteristic modifications of the photon spectrum, including energy-edge shifts, the appearance of higher-order harmonics, and spectral broadening. The second process is non-linear Breit–Wheeler pair production, where a high-energy photon converts into an electron–positron pair in the presence of the laser field. Measuring the pair-production probability as a function of laser intensity provides direct sensitivity to the non-perturbative structure of QED in strong fields and enables tests of theoretical predictions beyond perturbative power-law scaling. Together, these observables allow LUXE to probe strong-field QED over a wide range of parameters and to benchmark numerical and analytical tools used to describe strong-field processes. A detailed description of the experimental layout, detector systems, and simulation framework can be found in Refs.~\cite{Abramowicz:2021zja, LUXE:2023crk}

\subsection{Opportunities at Future Colliders}

While LUXE is optimized for electron–laser collisions at the European XFEL, strong-field effects are expected to play an increasingly important role at future accelerator facilities. As beam energies rise, the effective field strength experienced in particle–particle or particle–field interactions increases, pushing collider environments closer to regimes where strong-field QED phenomena become unavoidable.

\begin{figure}[!tbh]
    \centering
    \includegraphics[width=0.8\linewidth]{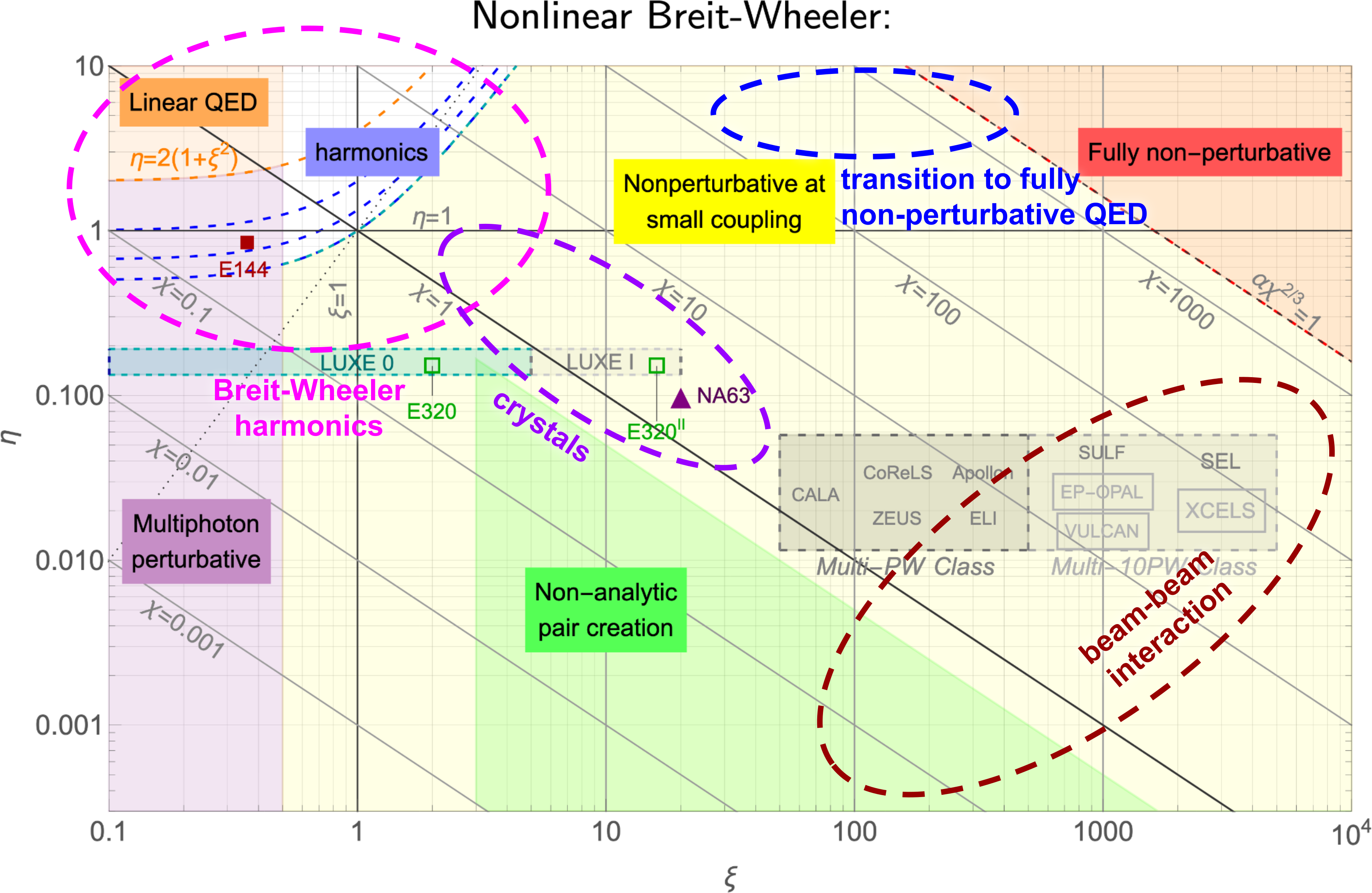}
    \caption{Overview of strong-field QED regimes relevant to non-linear Breit--Wheeler pair production accessible in different experimental environments as a function of the relevant interaction parameters. The figure highlights the parameter space that can be probed using electron–laser interactions, aligned crystals, and beam–beam effects. Future high-energy collider facilities extend the reach toward higher energies and intensities, building on the experimental program established by LUXE. Figure adapted from Ref.~\cite{Fedotov:2022ely}. }
    \label{fig:breitWheelerLandscapeFC}
\end{figure}

Several parts of the strong-field QED parameter space are of particular interest at future accelerator facilities, as indicated by the dashed ellipses in Figure~\ref{fig:breitWheelerLandscapeFC}. Compared to LUXE, beam--laser interactions at higher electron energies enable probing the harmonic structure of Breit--Wheeler pair production and provide access to the fully non-perturbative regime, in which perturbative QED breaks down and reliable theoretical predictions are currently lacking. Beam--beam interactions at lepton colliders give rise to beamstrahlung, and at the high energies foreseen for future facilities the associated electromagnetic fields can enter the strong-field regime, where non-linear QED effects may become relevant for beam dynamics and precision measurements. At present, no fully reliable simulation framework exists for beam--beam interactions at TeV-scale energies, highlighting the need to develop and benchmark improved simulation tools at facilities operating at lower energy. In addition, crystal-based experiments provide a complementary and experimentally established approach that can be integrated into future facilities and explored with higher energies, intensities, and improved control compared to existing setups.

\section{New-Physics Searches with Photons}

Strong-field interactions at accelerator facilities naturally produce intense beams of high-energy photons via the non-linear Compton process. Beyond serving as observable signatures of strong-field QED, these photons provide an attractive starting point for fixed-target searches for physics beyond the Standard Model of particle physics. In particular, beam-dump experiments offer a framework for searching for weakly coupled new particles over a broad mass range.

In a photon beam-dump experiment, a high-energy photon beam impinges on a dense target, where new particles can be produced via electromagnetic interactions with the nuclei of the dump material. A well-studied example is Primakoff production, in which photons convert into light scalar or pseudo-scalar states, such as axion-like particles (ALPs), in the electromagnetic field of the nucleus~\cite{Primakoff:1951iae}. If sufficiently long-lived, these states can traverse the dump and decay downstream into visible final states, most notably photon pairs.

Beam-dump experiments using photons can offer advantages over approaches based on charged particle beams under suitable experimental conditions. In particular, photons couple directly to axion-like particles via the Primakoff interaction, avoiding the intermediate bremsstrahlung step required in charged-beam experiments. Moreover, when high-energy photon beams with well-defined spectra are available, the background can be strongly suppressed, leading to comparatively clean electromagnetic final states~\cite{Bai:2021gbm}. As a result, photon-induced beam-dump experiments provide a well-motivated and complementary approach to searches for weakly coupled new particles.

\subsection{LUXE-NPOD and Future Collider Opportunities}

The LUXE-NPOD concept builds directly on the photon fluxes available from strong-field interactions~\cite{Bai:2021gbm}. The high-energy photons are directed onto a short dump, followed by a decay volume and a high-granularity electromagnetic calorimeter to detect photon pairs from particle decays. In the configuration considered here, the tungsten dump has a length of 2~m, followed by a decay volume of 10~m and a downstream detector with a radial acceptance of 5~m. Assuming a background-free environment, sensitivity estimates can be derived from simple geometric acceptance and decay probability considerations.

Within this framework, several accelerator environments can be considered. The LUXE-NPOD case can be optimized for higher photon fluxes, which naturally extend the physics reach. At future electron–positron colliders, photons from existing inverse Compton sources can be used (e.g.\ the FCCee intensity control) or produced in strong-field processes in dedicated interaction regions or extraction lines (e.g.\ at the FCCee injector or the tune-up extraction line of an ILC-like facility). Although the realizations differ, the underlying scaling of sensitivity is governed primarily by the photon energy, photon flux, and geometric configuration.

\begin{figure}[!tbh]
    \centering
    \includegraphics[width=0.8\linewidth]{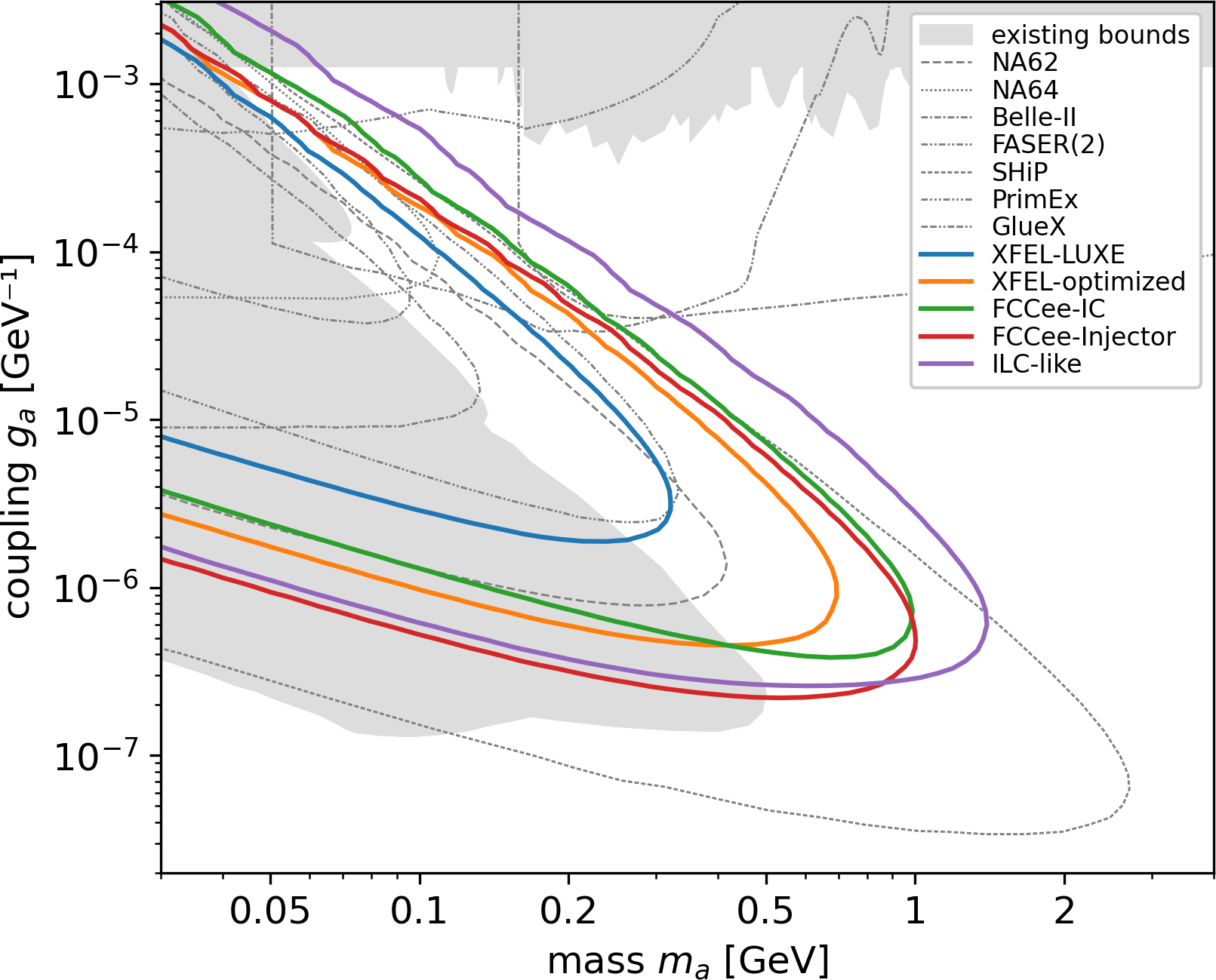}
    \caption{Projected sensitivity estimates at 95\% C.L. of photon-beam-dump searches for axion-like particles coupling to photons at different accelerator facilities, assuming one year of operation ($10^7$ seconds) in a background-free configuration. The projections are based on photon spectra and geometric acceptances characteristic of the considered setups and are intended to demonstrate the relative scaling with photon energy and flux. A detailed description of the underlying simulation framework and assumptions is given in Ref.~\cite{Schulthess:2025tct}. }
    \label{fig:phaseSpaceCoverage}
\end{figure}

The projected sensitivities can be compared with existing constraints and near-term experimental reach using the AxionLimits database, which provides a comprehensive and regularly updated overview of particle searches with axions and axion-like particles~\cite{AxionLimits}. As presented in Figure~\ref{fig:phaseSpaceCoverage}, the estimates of future collider infrastructures indicate that they could probe parts of the parameter space yet unexplored by existing constraints or projections from current experiments. Higher photon energies primarily extend the accessible mass range, while increased photon flux improves sensitivity to smaller couplings and enhances reach across the accessible mass spectrum. A detailed description of the simulation framework and extended parameter studies underlying these projections is presented in Ref.~\cite{Schulthess:2025tct}. In this broader context, the role of strong-field QED studies and photon-based new-physics searches at future linear and circular electron--positron colliders has been examined in recent community reports and conceptual studies~\cite{LinearColliderVision:2025hlt, Agapov:2928809}.

\section{Summary and Outlook}

Strong-field quantum electrodynamics provides access to a regime of QED in which non-perturbative effects become experimentally testable. The LUXE experiment is designed to explore this regime using electron–laser collisions in a controlled accelerator environment, enabling precision studies of non-linear QED processes and benchmarking theoretical and numerical descriptions of strong-field interactions.

Beyond these measurements, LUXE provides a framework that can be extended to future collider facilities, where strong electromagnetic fields are expected to play an increasing role. The high-energy photon beams produced in strong-field interactions also enable searches for new physics in photon beam-dump experiments. Sensitivity estimates indicate that future accelerator infrastructures could explore regions of parameter space for light, weakly coupled particles that extend those currently probed by existing experiments.

\backmatter

\bmhead{Acknowledgements}
The author thanks F.~Meloni for valuable discussions related to this work. This research was inspired by and built on work within the LUXE Collaboration, whose scientific environment and resources are gratefully acknowledged. This work was supported by the Swiss National Science Foundation under grants no. 214492 and 230596. AI-assisted tools, including OpenAI ChatGPT and Writefull, were used to improve phrasing and clarity in the preparation of this manuscript.

\bibliography{ref}

\end{document}